\documentstyle[preprint,prl,aps]{revtex}
\tightenlines
\begin{document}
\draft
\title{Oscillation modes of two-dimensional nanostructures within
the time-dependent local-spin-density approximation}
\author{Antonio Puente and Lloren\c{c} Serra}
\address{Departament de F\'{\i}sica, Universitat de les Illes Balears,
E-07071 Palma de Mallorca, Spain}
\date{April 26, 1999}
\maketitle
\begin{abstract}
We apply the time-dependent local-spin-density approximation 
as general theory to describe ground states and spin-density oscillations 
in the linear response regime of two-dimensional 
nanostructures of arbitrary shape. For this purpose, a frequency analysis 
of the simulated real-time evolution is performed. The effect on the response 
of the recently proposed spin-density waves in the ground
state of certain parabolic quantum dots is considered. They lead to 
the prediction of a new class of excitations, soft spin-twist modes, with 
energies well below that of the spin dipole oscillation.
\end{abstract}

Recent advances in semiconductor technology nowadays allow the 
fabrication of nanostructures with many different shapes. In these 
systems the electrons, which are laterally confined at the semiconductor 
boundary, form a two-dimensional quantum dot with a shape 
which, to a certain extent, follows that of the nanostructure. This 
opens up the exciting possibility to produce and study an enormous variety of
quantum dots, or {\em artificial atoms} as they are often called. 
For instance, it has been shown that the electronic 
structure in the small vertical quantum dots of Ref.\ \onlinecite{Tar97} 
is given by the successive filling of shells obeying Hund's rules as in atoms.
Very relevant information about electronic excitations in quantum dots 
is also presently obtained from 
sophisticated far-infrared absorption \cite{Dem90}
and light scattering experiments \cite{Sch96}.
  
Up to now, the great majority of experimental and theoretical efforts 
were focussed on quantum dots with circular symmetry.
Many of the properties of circular dots are well reproduced by 
considering the electrons as confined by a parabolic potential, or by a 
simple jellium disk. To treat the electronic interactions, besides 
exact diagonalization for very small 
dots \cite{Yan93}, the most succesful approaches
have been mean field theories like Hartree-Fock (HF) \cite{Gud91} 
and density functional 
in the local-spin-density approximation \cite{Kos97,Ser97} (LSDA) . 

The latter ones have been 
extended using the random-phase approximation (RPA) to analyze collective 
excitations \cite{Gud91,Ser99}. To our knowledge, 
all theoretical approaches addressing collective excitations in 2d quantum 
dots are limited from the start by the circular symmetry assumption. 
In this Letter we show how LSDA can describe both ground state
and linear response of 2d quantum dots of arbitrary shape by using,
respectively, energy minimization and real time simulation of the 
spin-density oscillations as basic principles. 
We will show how from the response 
frequencies in the different channels (density, spin and free responses)
it is possible to gain information about the system deformation in 
a quantitative way. Besides, we will also analyze the effect on the 
response of the recently proposed spin-density waves in the ground state 
of particular parabolic quantum dots. Static spin-density waves could 
manifest their existence by means of soft spin-twist modes, at energies 
well below that of dipole spin oscillation. 

Several authors have recently addressed the problem of describing quantum
dot ground states within LSDA. In particular, in Ref.\ \onlinecite{Kos97}
the single particle Kohn-Sham equations for electrons in a parabolic 
potential were solved avoiding any symmetry restriction by using a
plane-wave basis. We use here the same LSDA functional 
of Ref.\ \onlinecite{Kos97}, i.e., the local functional
based on the von Barth-Hedin \cite{vonB72} interpolation of the 
Tanatar-Ceperley \cite{Tan89}
results for the non-polarized and fully polarized 2d electron gas. However, 
we employ a different technique, based on the discretization 
of the $xy$ plane in a grid of uniformly spaced points. 
For each spin ($\eta=\uparrow,\downarrow$) the Kohn-Sham equations read
\begin{eqnarray}
\label{eq1}
\left[
-{1\over 2}\nabla^2 + v^{({\em conf})}({\bf r}) + 
v^{(H)}({\bf r}) + v^{(xc)}_\eta({\bf r})
\right]
\varphi_{i\eta}({\bf r}) && \\ \nonumber
= \epsilon_{i\eta} \varphi_{i\eta}({\bf r}) &&
\; ,
\end{eqnarray} 
where $v^{({\em conf})}({\bf r})$ and 
$v^{(H)}({\bf r})=\int{d{\bf r}' \rho({\bf r}')/{|{\bf r}-{\bf r}'|}}$
are, respectively, the confining and Hartree potentials. The 
exchange-correlation contributions are obtained from the 
local energy density ${\cal E}_{xc}(\rho,m)$ by
\begin{equation}
v^{(xc)}_\eta({\bf r})=
{\partial\over\partial\rho_\eta}{\cal E}_{xc}(\rho,m)\; .
\end{equation}
We have defined 
total density and magnetization, 
in terms of the spin densities 
$\rho_\eta({\bf r})=\sum_i{|\varphi_{i\eta}({\bf r})|^2}$,
as 
$\rho=\rho_\uparrow+\rho_\downarrow$ and
$m=\rho_\uparrow-\rho_\downarrow$, respectively.

As a test of the numerical code using the $xy$ grid we have
checked that for a circular dot (namely, the parabolic one
confined by ${1\over 2}m\omega_0^2 r^2$, with $r$ the radial 
coordinate, $\omega_0=0.25$~H$^*$ \cite{not1} and $N=20$
electrons) we find the 
same solution that is obtained by solving only the radial
equation and imposing $e^{-i\ell_i\theta}$ as the 
angular part of the single particle 
wave functions. Next, we have considered different confining 
geometries for dots with $N=20$ electrons. In particular, we present 
here results for a deformed parabola 
${1\over2}m(\omega_x^2 x^2+\omega_y^2 y^2)$
with $\omega_y=0.75\,\omega_x=0.22$H$^*$; for a square jellium with 
$r_s=1.51$, that 
corresponds to side length $L=11.96 a_B^*$; and for a rectangular 
jellium with the same $r_s$ and sides $L_y=0.75\,L_x=10.37 a_B^*$. 
A more systematic investigation of ground states for different sizes 
and different confining geometries is left for future work. Here our 
aim is mainly to show the feasibility of the method and to concentrate 
on the spin-density oscillations.

The previously mentioned ground states are shown in Fig.\ 1.
In the deformed parabola, the density shows an ellipsoidal shape
with an aspect ratio similar to $\omega_y/\omega_x$. Three rows of local 
maxima can be seen in the inner part of the dot, aligned with the long
axis. For the square jellium we obtain a rather abrupt  electron 
density, with maxima at the corners and four additional inner maxima
following the square symmetry. A similar structure is seen for the 
rectangle. Quite interestingly, while for the 
deformed parabola the magnetization vanishes everywhere,
for the square and rectangle there is a magnetization wave in the ground
state. The amplitude of this wave is approximately 15\% and 25\% of the 
maximum density, respectively.
This finding is similar to the spin density waves predicted by 
Ref.\ \onlinecite{Kos97} in some circular parabolic dots.

The description of spin-density oscillations in quantum dots has 
raised great interest, mainly due to the manifestation of these modes
in far-infrared absorption and in Raman scattering 
experiments \cite{Dem90,Sch96}. 
We refer
here to general spin-density oscillations. When both spin components
oscillate in phase they produce density modes and when they are out of 
phase, spin modes. In circularly symmetric dots, density modes have been 
studied using the Hartree \cite{Broi90}
and Hartree-Fock \cite{Gud91} methods. More recently 
the LSDA to density functional theory has been used in circular dots to 
describe 
density and spin channels \cite{Ser99}, taking into account the coupling
between both.
All these methods are based on the perturbative treatment of the response
by diagonalization of the residual interaction within a 
space of particle-hole excitations. They share as essential ingredient
the angular momentum selection rules given by the circular symmetry.
In fact these methods use the well known RPA, based on different ground state
theories.

To describe spin-density oscillations in an arbitrary structure the
RPA approach becomes practically unfeasible because of the enormous
dimension of the matrices. This is due to the lack 
of symmetry, which forces to deal with matrices
$\chi(x,y,x',y';\omega)$ in the the formal RPA equation
$\chi = \chi^{(0)} V_{ph} \chi$, 
where $\chi^{(0)}$ is the independent particle correlation function
and $V_{ph}$ is the residual particle-hole interaction. The calculation
is also complicated due to 
the breaking of degeneracies with deformation, that greatly increases
the number of different particle-hole pairs contributing to 
$\chi^{(0)}$. An alternative approach that permits to overcome these 
problems is based on real-time methods. These originate in fact 
from the time-dependent HF theory, and have been applied with success
to nuclear \cite{Bon76} and to cluster 
physics \cite{Yab96}. In what follows we briefly 
comment this approach and show how it applyies to 2d nanostructures.

In the small amplitude limit, it is well known that both 
real time (TDHF, TDLSDA) and RPA methods based on the corresponding 
ground states become equivalent. 

We have performed TDLSDA calculations 
by integrating the time-dependent Kohn-Sham equations
\begin{equation}
i{\partial\over\partial t} \varphi_{i\eta}({\bf r},t) =
h_\eta[\rho,m]\, \varphi_{i\eta}({\bf r},t)\;,  
\label{eq5}
\end{equation}
where $h_\eta$ is given by the square bracket in Eq.\ (\ref{eq1}).
We have integrated Eq.\ (\ref{eq5}) using the Crank-Nicholson 
algorithm (for time step $n$ to $n+1$)
\begin{equation}
\left( 1+{i\Delta{t}\over 2} h_\eta^{(n+1)}\right) 
\varphi_{i\eta}^{(n+1)} =
\left( 1-{i\Delta{t}\over 2} h_\eta^{(n)}\right) 
\varphi_{i\eta}^{(n)}\;.
\label{eq5b}
\end{equation}
This is in an implicit problem for $\varphi^{(n+1)}$, 
since $h_\eta^{(n+1)}$ depends on the 
orbitals through the density and magnetization. In practice this 
forces to proceed by iteration: with $h_\eta$ from the previous time 
step, a first guess of the new wave functions is obtained solving 
(\ref{eq5b}).
These are then used to build a new $h_\eta$ and restart iteration. 
Using a rather small time step it is enough to make a double solution
for each time step. 
If $h_\eta$ were constant in time, the algorithm would be 
exactly unitary. 
We have found that with small $\Delta{t}$ norm conservation is fullfiled
with excellent accuracy. 

In order to excite the oscillation modes of the system an initial 
perturbation of the wave functions is needed. Physically, this 
corresponds for instance to the interaction with a short laser pulse 
or with an appropriate projectile. In the calculation, it can be mimicked 
simply by a rigid translation of the wave functions by means of the 
operator 
${\cal T}({\bf a}_\sigma)=e^{-{\bf a}_\sigma\cdot\nabla}$
or by 
an initial impulse with 
${\Pi}({\bf q}_\sigma)=e^{-i{\bf q}_\sigma\cdot{\bf r}}$.
When either ${\bf a}_\sigma$ or ${\bf q}_\sigma$ are small these 
perturbations induce dipole oscillations predominantly and the 
system's response is restricted to the linear regime. Total density 
and spin modes are obtained with the rigid translations  
${\bf a}_\uparrow={\bf a}_\downarrow={\bf a}$, and 
${\bf a}_\uparrow=-{\bf a}_\downarrow={\bf a}$, respectively. With the 
impulse initial conditions these are: 
${\bf q}_\uparrow={\bf q}_\downarrow={\bf q}$ (density),
${\bf q}_\uparrow=-{\bf q}_\downarrow={\bf q}$ (spin).
After the initial perturbation we keep track of the time dependent 
dipole moments $d_t$: $\langle {\bf e}\cdot{\bf r}\rangle_t$ for density 
modes and $\langle {\bf e}\cdot{\bf r}\sigma_z\rangle_t$ for spin modes.
Here ${\bf e}$ corresponds to the direction of the initial 
perturbation given by ${\bf a}$ or ${\bf q}$.

A frequency analysis of the dipole signal $d_t$ gives the response frequencies 
of the system. Fourier tranform methods can be used for this purpose. 
However, we have found more efficient a method of direct peak fitting to the 
simulated signal. We perform a least squares minimization
of $\chi^2=\sum_t{(d_t-D(t))^2}$, where $D(t)$ is given by
\begin{equation}
D(t)=\sum_{n=1}^N{A_n\cos(\omega_n t)+B_n\sin(\omega_n t)}\; .
\label{eq6}
\end{equation}
In Eq.\ (\ref{eq6}) a fixed number of frequencies is assumed. The minimization
yields the amplitudes $A_n$, $B_n$ and frequencies $\omega_n$. Of course, 
it must be checked that the number $N$ of frequencies is large enough 
to provide a good reproduction $D(t)$ of the time series $d_t$ and  
that convergence with increasing $N$ has been reached. 
From the fitted $D(t)$ we obtain $D(\omega)$ as a discrete set 
of Dirac delta functions.
In practice, these are smoothed into Lorentzians and 
the response strength is obtained as $S(\omega)=|D(\omega)|$, or the
power spectrum as ${\cal P}(\omega)=|D(\omega)|^2$.

We have performed the response calculation in real time for the same dots 
with 20 electrons discussed above.
An excellent agreement with the RPA calculation was obtained for 
the circular parabola.
Figure 2 shows the three responses for the deformed parabola, 
for which the magnetization $m({\bf r})$ is vanishingly small. 
The density response has only two peaks that coincide with the 
parabola frequencies $\omega_x=0.29$
and $\omega_y=0.22$. This is showing that TDLSDA satisfies 
the generalized Kohn's theorem for a deformed parabola.
Since the density dipole operator only couples to the center of mass motion, 
absorption in this channel can only take place exactly at the frequencies 
$\omega_x$ and $\omega_y$.
The situation is completely different for 
the free and spin responses. They are rather 
fragmented and lie at lower energies. 
Peaks associated to oscillation along different axis are shown 
in different line types. 
The free response corresponds to keep the 
effective confining potential in (\ref{eq5}),
$v^{({\em conf})}+v^{(H)}+v_\eta^{(xc)}$, fixed to its static value. 
It thus models the oscillation of non interacting particles in the 
static mean field.
By comparing free, density and spin responses we see
the different nature of the residual interaction in both channels: weakly 
attractive in the spin response and repulsive in the density one.
Figure 2 also shows the simulated time series in each case, as well as the 
fitted signal which on the plot scale superposes to the simulated one.

Figure 3 shows the corresponding results for the $N=20$ electrons 
dots in the square and rectangular jellium. The density response 
of the square is still characterized by a very dominant peak that, 
nevertheless,  is slightly fragmented. The free and spin responses are more 
fragmented. Quite interestingly, the spin response of the square shows
four groups of peaks with an almost constant separation of 0.05H$^*$.
In the rectangle case, the density response is clearly showing   
that the oscillation frequencies are different
in $x$ and $y$ directions. The same fact is an additional source 
of fragmentation for the free and spin channels.

In circular parabolic dots LSDA predicts spontaneous symmetry breaking
ground states, of the type of spin density waves, for particular 
sizes \cite{Kos97,Hir99}.
This LSDA spin density wave is more pronounced in quasi 
one-dimensional rings \cite{Rei99}.
In Ref.\ \onlinecite{Yan99} similar spin density waves, as well as 
Wigner crystallized ground states, have been very recently reported in 
circular parabolic dots using an unrestricted Hartree-Fock approach. 
There is, however, a continuing discussion about the 
interpretation and possible relevance of these 
states \cite{Kos97,Hir99,Yan99}. We present in the rest of this Letter the 
TDLSDA result for the linear oscillations of a spin-density-wave 
ground state in a circular dot.
 
Finally, we have analyzed the oscillations of a dot with $N=24$ electrons 
in circular parabolic confinement ($\omega_0=0.24$~H$^*$), 
for which there is a static spin-density 
wave in the LSDA ground state \cite{Kos97}. We have found that the 
dipole oscillations of spin densities are quite similar to those obtained from
a fully circular model \cite{Ser99}. However, the spin density wave
can sustain a new type 
of oscillation. It is given by an alternating rotation of both spin densities
in opposite directions, i.e., a spin twist of the static wave
excited with the rotation operator ${\cal R}(\theta_\sigma)$, 
with $\theta_\uparrow=-\theta_\downarrow$ being opposite rotation
angles for each spin.
The frequency of 
this mode is obtained analyzing the time evolution of the circular 
currents that appear after an initial rotation 
with ${\cal R}(\theta_\sigma)$. Figure 4 shows the strength 
of the spin-twist mode, in comparison with the normal spin dipole mode.
Spin-twist modes are very soft, with energy well below the spin dipole one.
They could signal the existence of static spin density waves
in circular dots \cite{not2}.
The spin-twist frequency is reflecting the curvature of the energy 
minimum corresponding to the symmetry broken ground state with respect to
the circular one. We expect that circular systems having strong spin 
density waves in their ground states will also exhibit enhanced spin-twist 
modes. This happens in circular parabolic dots with increasing $r_s$ 
for a fixed $N$ \cite{Kos97} (the $N$-systematics with fixed $r_s$ is less
clear) or in quasi one dimensional rings \cite{Rei99}.

In conclusion, we have shown that TDLSDA can be applied to obtain 
the oscillation frequencies of nanostructures with arbitrary shape.
It leads to the prediction of soft spin-twist modes in dots with circular 
parabolic confining and having static spin-density waves in their 
ground state.

This work was performed under
Grant No.\ PB95-0492 from CICYT, Spain.

\begin{figure}[h]
\caption{Two-dimensional surface and contour-line plot of the 
ground state density (two upper rows) and magnetization (two lower rows)
of dots with $N=20$ electrons in a deformed parabola, a square jellium 
and a rectangular jellium, respectively. The magnetization of the deformed
parabola vanishes everywhere. See text.}
\end{figure}

\begin{figure}[h]
\caption{Simulated and fitted time series (left), and strength functions
in arbitrary units
(right) for the deformed parabola corresponding to free, density
and spin excitations. See text.}
\end{figure}

\begin{figure}[h]
\caption{Strength functions in arbitrary units for a square and rectangular
jellium. Free, density and spin responses
are shown. For the square, only the total spectrum is 
displayed since $x$ and $y$ oscillations are essentially degenerate.}
\end{figure}

\begin{figure}[h]
\caption{First row shows the spin-density wave in a parabolic dot of $N=24$
electrons as a 2d surface and contour plot. Middle panel 
shows the spin dipole strength, while lower one shows the strength 
of the spin-twist mode in arbitrary units.} 
\end{figure}


\begin{references}
\bibitem{not1} We use effective atomic units corresponding 
to GaAs values, i.e., ${\rm H}^*\approx 12 {\rm meV}$, 
${\rm a}^*_B\approx 98 {\rm\AA}$ and 
$\tau^* = 55$~fs as energy, length and time units.
\bibitem{not2} Differently to the circularly symmetric confining 
potential, in non circular confinements (deformed harmonic oscillator,
square, rectangle, \dots) spin-twist modes will exist irrespective 
of the existence of spin-density-waves in their ground state. This is 
because in deformed systems the circular symmetry is obviously broken.
\end{references}
\end{document}